\documentclass[journal]{IEEEtran}
\usepackage{bm}
\usepackage[colorlinks,linkcolor=.,anchorcolor=., citecolor=.,filecolor=.,menucolor=., runcolor=., urlcolor=blue]{hyperref}
\usepackage{textgreek}
\usepackage{float}
\usepackage{placeins}
\usepackage{graphicx}
\usepackage{graphicx}
\usepackage{textcomp}
\usepackage{xcolor}
\usepackage{multirow}
\usepackage{caption}
\usepackage{subcaption}
\usepackage{footnote}
\makesavenoteenv{tabular}
\makesavenoteenv{table}
\usepackage{cite}
\usepackage{amsmath,amssymb,amsfonts}
\usepackage{algorithmic}
\usepackage{graphicx}
\usepackage{textcomp}
\usepackage{graphics}
\begin{document}
\fboxrule=0pt
\bstctlcite{IEEEexample:BSTcontrol}
\hyphenpenalty=10000
\exhyphenpenalty=10000
%
\title{Wafer-Level 1/f Noise Characterization of Scaled P-Type Nanosheets and Comparison with Planar HKMG Transistors}
%
%
%

\author{Ruben Asanovski\textsuperscript{1,*}, Anabela Veloso\textsuperscript{1}, Ben Kaczer\textsuperscript{1}, Naoto Horiguchi\textsuperscript{1}, and Jacopo Franco\textsuperscript{1} 
\\ \ \\
{\small \textsuperscript{1} imec, Leuven, Belgium\\
\textsuperscript{*} E-mail: \href{mailto:ruben.asanovski@imec.be}{ruben.asanovski@imec.be}  
}}
%
%
\markboth{}%
{Asanovski \MakeLowercase{\textit{et al.}}: Impact of Gate Metal Work Function on 1/f Noise in RMG MOSFETs}
%



\maketitle

\begin{abstract}
We present a comprehensive wafer-level 1/f noise study of scaled Gate-All-Around (GAA) p-type nanosheet transistors and benchmark them against large-area planar High-$\kappa$ Metal Gate (HKMG) devices to assess the impact of device architecture on reliability.  By statistically analyzing noise data from 188 nanosheets, we extract the effective trap density in the dielectric and compare it to that of planar transistors with identical gate stack, comparable processing thermal budgets, and no specific reliability anneals. The effective trap density extracted from 1/f noise is comparable in planar and nanosheet architectures. This indicates that transitioning to GAA nanosheets does not increase noise, which remains governed by gate stack quality.
\end{abstract}

\begin{IEEEkeywords}
Nanosheet, 1/f noise, reliability, gate stack
\end{IEEEkeywords}

%
\IEEEpeerreviewmaketitle

\section{Introduction}
GAA nanosheet transistors are a leading candidate for advanced CMOS nodes due to their superior electrostatic control and scalability \cite{Colinge90,Loubet17,Yeap24}. Assessing reliability in scaled technologies is challenging, as conventional methods such as C(G)-V dispersion \cite{Nicollian02} and charge pumping \cite{Groeseneken84}, are not easily applicable to nanoscale floating body devices. In contrast, 1/f noise offers a non-destructive, wafer-level method to evaluate gate stack quality \cite{Christensson68,Kirton89,Jayaraman89,Giusi06,Ghibaudo91}. Moreover, 1/f noise correlates well with other reliability metrics, particularly Bias Temperature Instability (BTI) \cite{Asanovski2023_SSE,Asanovski2024,Asanovski2025,Goes14,Ding16} since both mechanisms mainly originate from stochastic charge trapping and de-trapping at pre-existing defects in the gate dielectric and interface.
\par In this study, we analyze the noise of scaled p-type nanosheet transistors and benchmark it against planar HKMG devices with identical gate stacks and comparable thermal processing. By analyzing noise data from a statistically significant number of nanosheet transistors and benchmarking against planar devices, we assess whether the GAA nanosheet architecture introduces any reliability penalty. 
\section{Measurement setups}
\begin{figure}[!tb]
\centering
\includegraphics[width=0.45\textwidth]{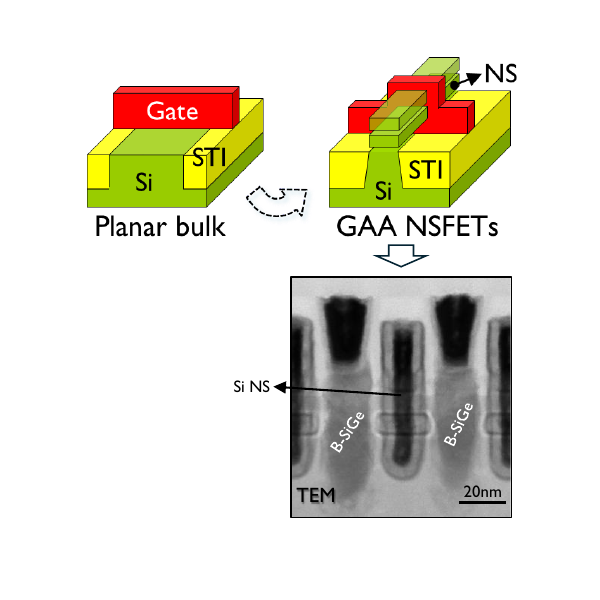}
\caption{Schematic comparison of planar bulk and GAA nanosheet FET architectures alongside a Transmission Electron Microscopy (TEM) image of the nanosheet transistor along the channel fabricated in imec.}
\label{fig:TEM}
\end{figure}

\begin{figure}[!tb]
\centering
\includegraphics[width=0.3\textwidth]{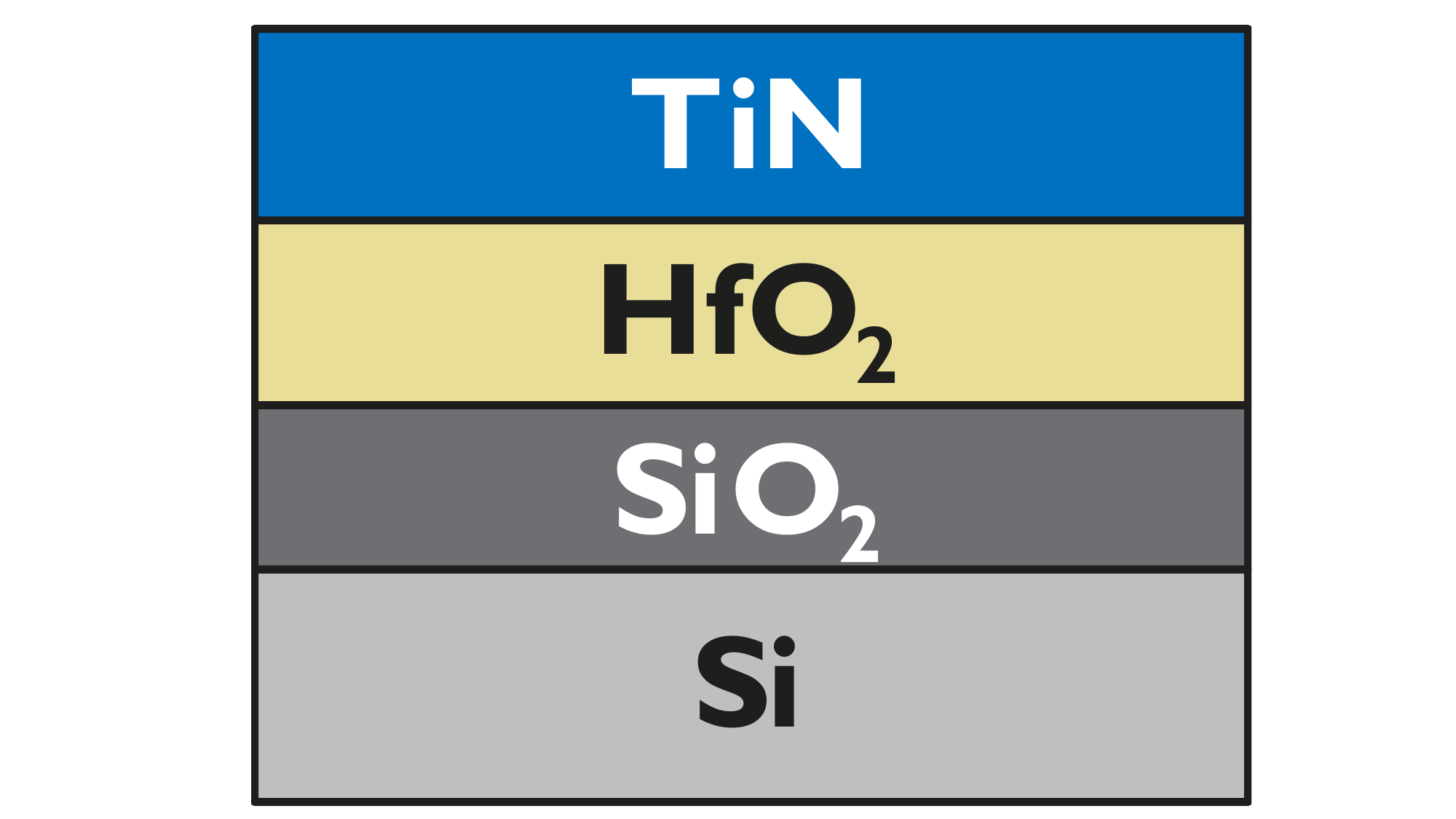}
\caption{Schematic of the HKMG stack employed in both planar and nanosheet pMOSFETs.}
\label{fig:GateStack}
\end{figure}

We measured scaled p-type nanosheet transistors (see Fig.~\ref{fig:TEM}) fabricated on 300 mm wafers \cite{Sarkar2025}, featuring a tight 48 nm Contacted Gate Pitch (CGP). The nanosheets feature a nominal Equivalent Oxide Thickness (EOT) of ~1 nm, gate width ($W$) of 17 nm, gate height ($H$) of 6.5 nm, and gate length ($L$) of 19 nm (as estimated from TEM images). The effective width of the device ($W_{\textrm{eff}}$) is 2($W$+$H$)=47 nm. The devices employ a standard high-$\kappa$ metal gate (HKMG) stack with TiN as the gate electrode (Fig.~\ref{fig:GateStack}). For benchmarking, we also characterized planar HKMG pMOSFETs with large dimensions ($W{=}L{=}1\ \textrm{\textmu m}$), featuring the same gate stack and same Replacement Metal Gate (RMG) stack thermal budget \cite{Asanovski2024,Asanovski2025,Franco23,Franco24,Vici24}. No dedicated reliability anneal was included in the RMG flow; only a conventional forming gas sintering anneal was performed at the end of the flow to passivate interface states.
\par The measurements consisted of $I_{\textrm{D}}-V_{\textrm{GS}}$ sweeps and 1/f noise in the linear region of operation ($V_{\textrm{DS}}$=-50~mV) at various drain currents (modulated via $V_{\textrm{GS}}$) and a temperature of $25^\circ\mathrm{C}$. To ensure statistical relevance, we measured 94 nanosheet devices, each with two adjacent sheets connected in parallel, for a total of 188 nanosheets across multiple dies. On the other hand, we measured only 12 planar devices for the comparison due to their large area. All measurements were done using a Keysight B1500 semiconductor parameter analyzer, in combination with a Keysight E4727B low-frequency noise analyzer. Noise measurements were made in the 10 Hz to 1 kHz frequency range to minimize acquisition time.
\section{Results and discussion}
\begin{figure}[!tb]
\centering
\subfloat[]{\includegraphics[width=0.24\textwidth]{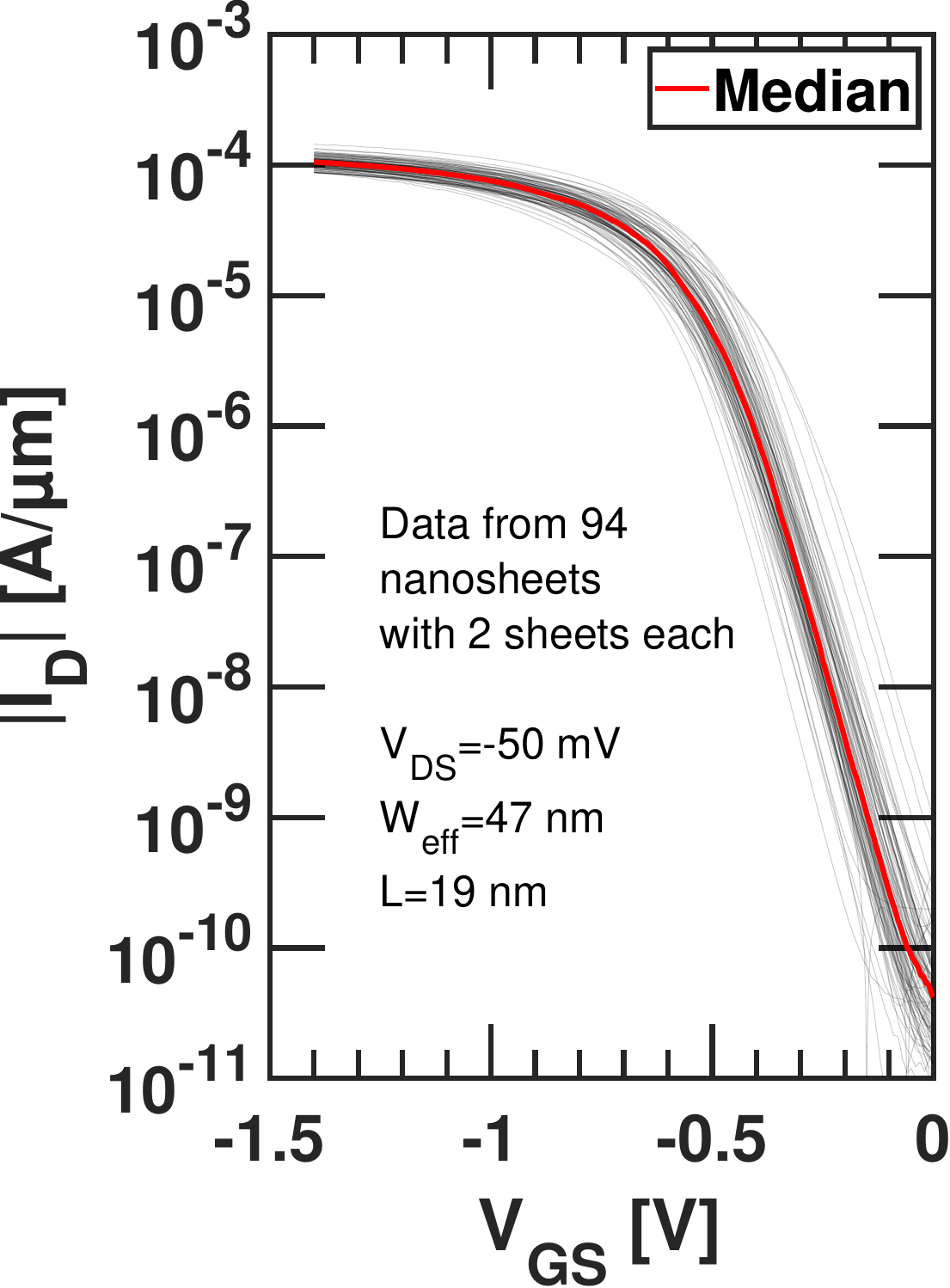}\label{fig:IdVg_NS_ALL}}  
\subfloat[]{\includegraphics[width=0.24\textwidth]{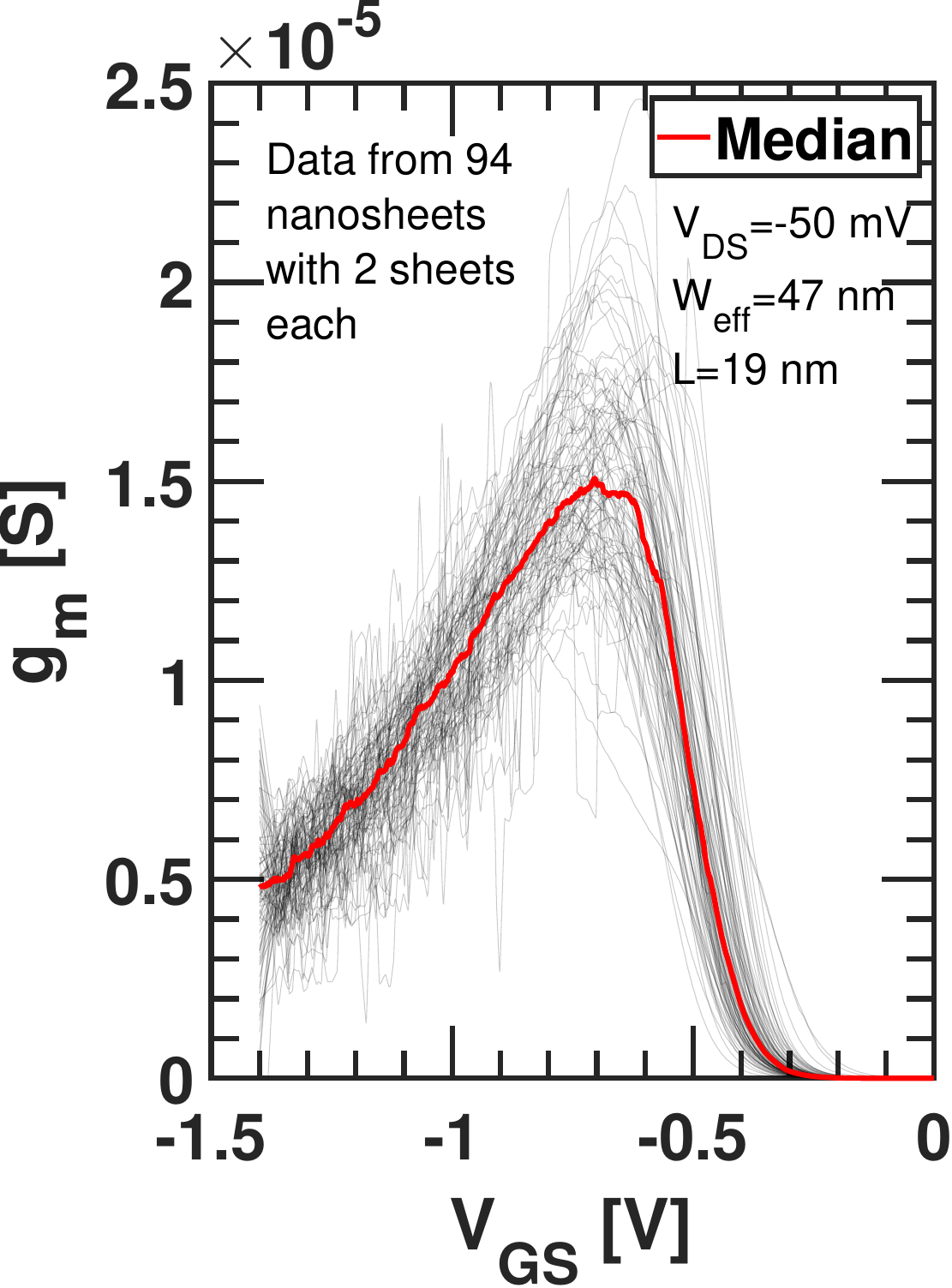}\label{fig:gm_NS_ALL}}
\caption{(a) $I_{\textrm{D}}-V_{\textrm{GS}}$ and (b) $g_{\textrm{m}}-V_{\textrm{GS}}$ in the linear region for 94 nanosheet pMOSFETs considered in this study. The median curve is highlighted in red.}
\label{fig:DC_NS_ALL}
\end{figure}

\begin{figure}[!b]
\centering
\subfloat[]{\includegraphics[width=0.24\textwidth]{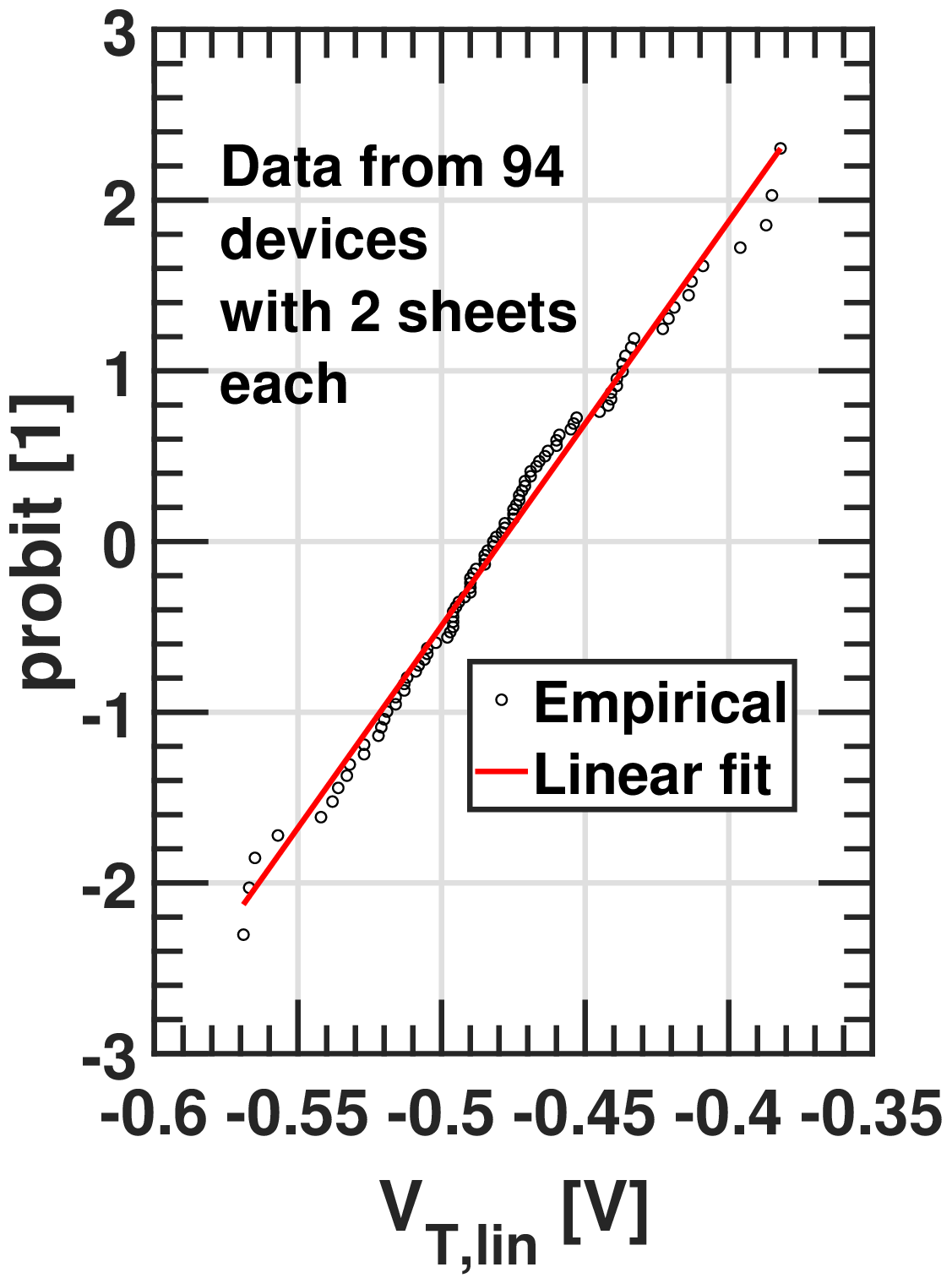}\label{fig:probit_Vt}}  
\subfloat[]{\includegraphics[width=0.24\textwidth]{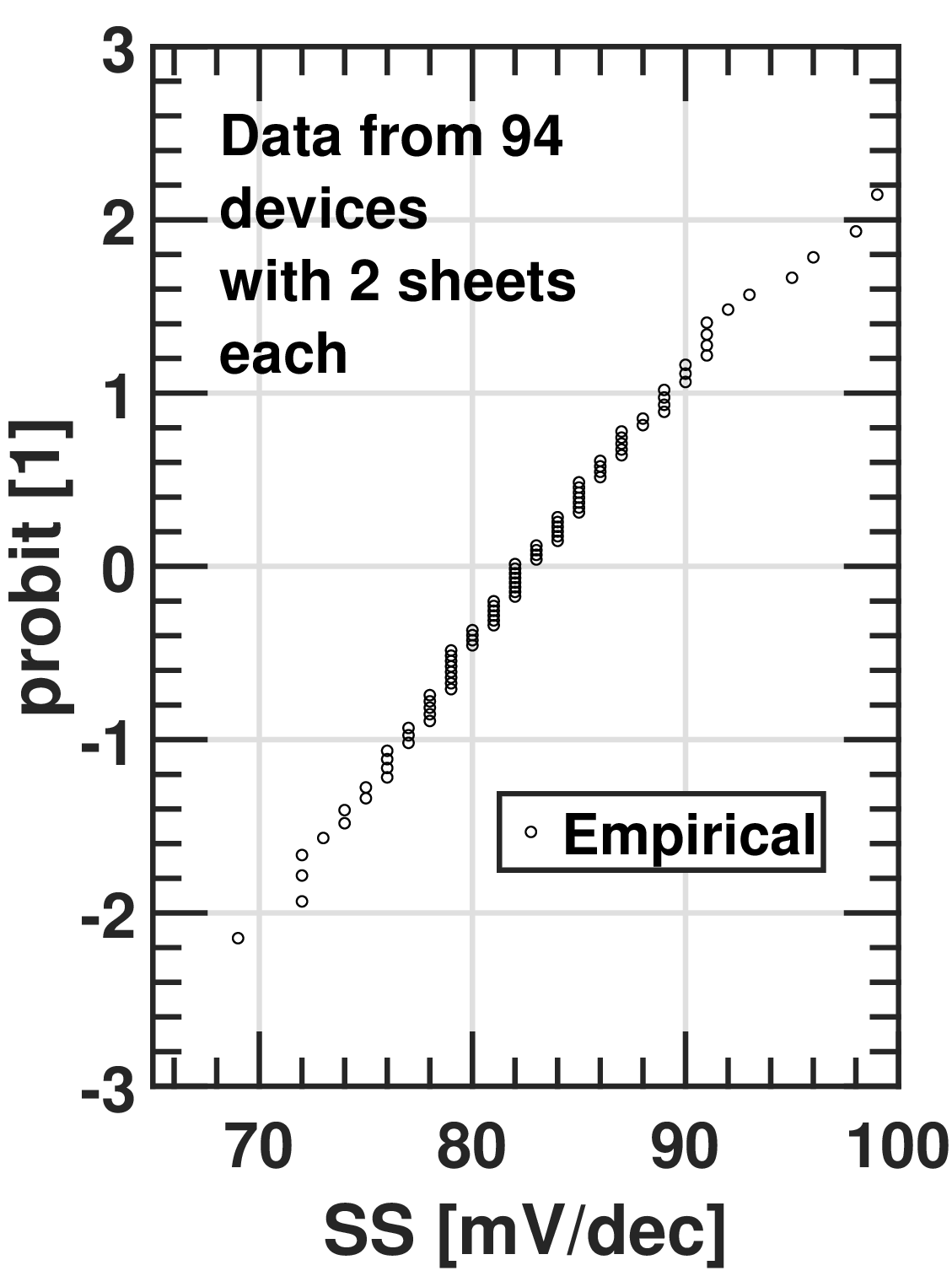}\label{fig:probit_SS}}
\caption{Probit plots of the CDF for (a) $V_{\mathrm{T}}$ and (b) SS of the 94 nanosheet pMOSFETs. The linearity of the $V_{\mathrm{T}}$ plot confirms a normal distribution.}
\label{fig:probit_DC}
\end{figure}

The $I_{\textrm{D}}-V_{\textrm{GS}}$ and $g_{\textrm{m}}-V_{\textrm{GS}}$ of the 94 nanosheets in the linear region (Fig.~\ref{fig:DC_NS_ALL}) show good device uniformity, with the median curve highlighted in red. The Cumulative Distribution Functions (CDFs) of threshold voltage ($V_{\mathrm{T}}$) and subthreshold swing (SS) are plotted in probit scale in Fig.~\ref{fig:probit_DC}. The $V_{\mathrm{T}}$ CDF exhibits a normal distribution, as confirmed by the linearity of the probit plot. Although the SS values in Fig.~\ref{fig:probit_SS} appear normally distributed, a Gaussian distribution is not physically valid because it would predict SS values below the Boltzmann limit \cite{Beckers2020}.
\par
\begin{figure}[!b]
\centering
\includegraphics[width=0.24\textwidth]{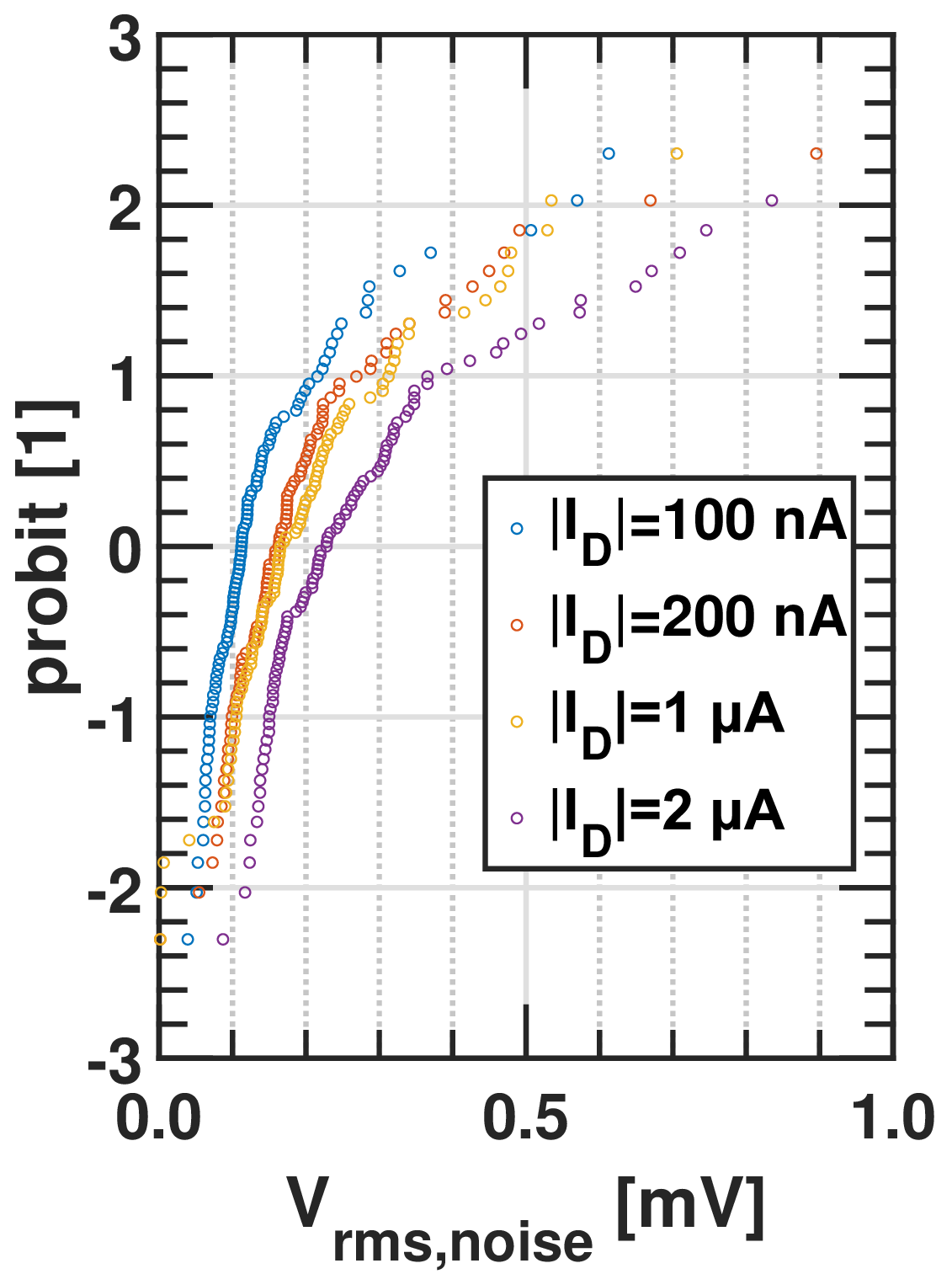}
\caption{Probit plot of the CDF for the integrated noise power $V_{\mathrm{rms,noise}}$ (Eq.~\ref{eq:Vrms}) for 94 nanosheet pMOSFETs at different $I_{\mathrm{D}}$ values.}
\label{fig:probit_Noise}
\end{figure}

\begin{figure}[!tb]
\centering
\subfloat[]{\includegraphics[width=0.45\textwidth]{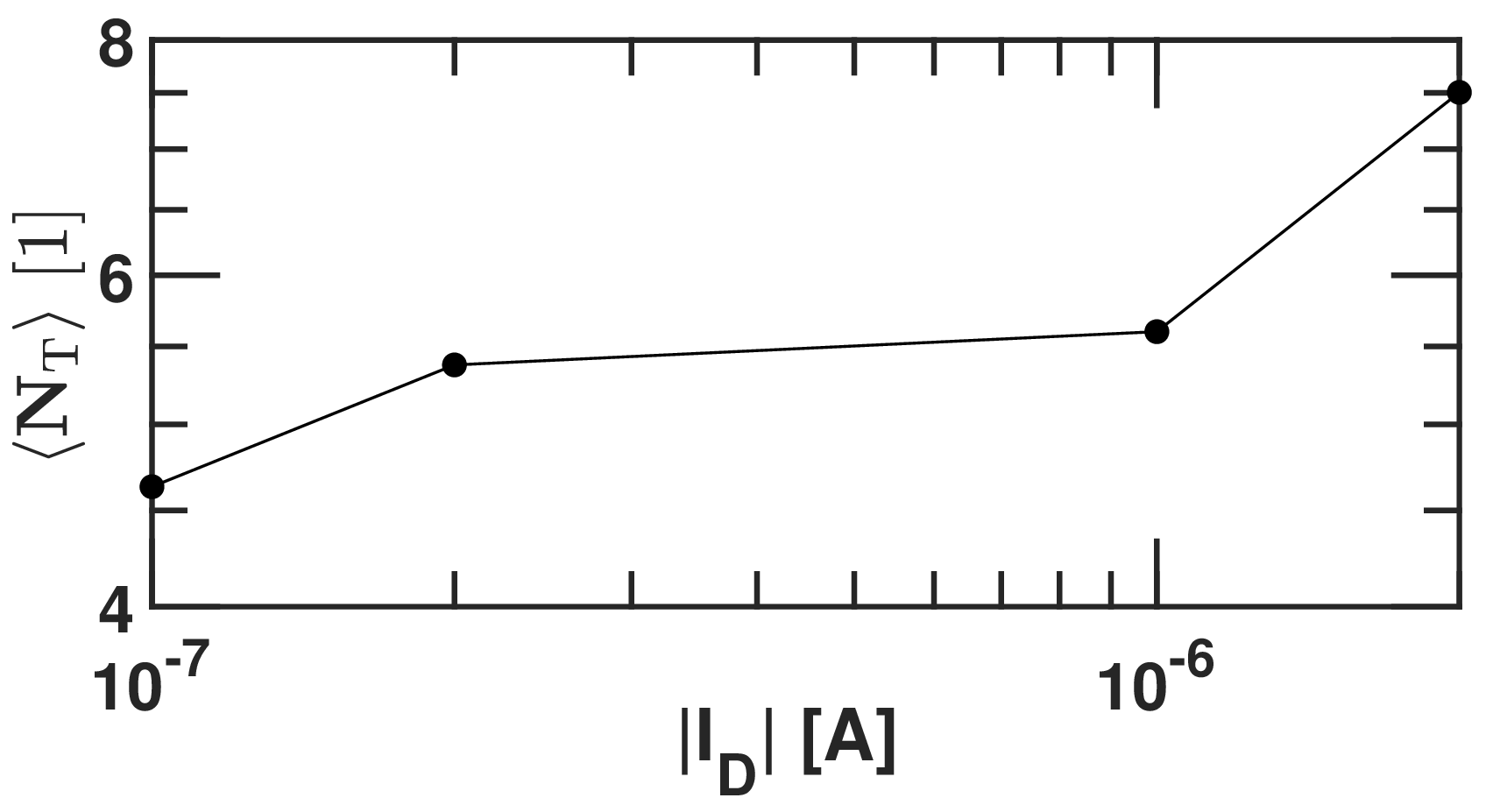}\label{subfig:NtId}}\\
\subfloat[]{\includegraphics[width=0.45\textwidth]{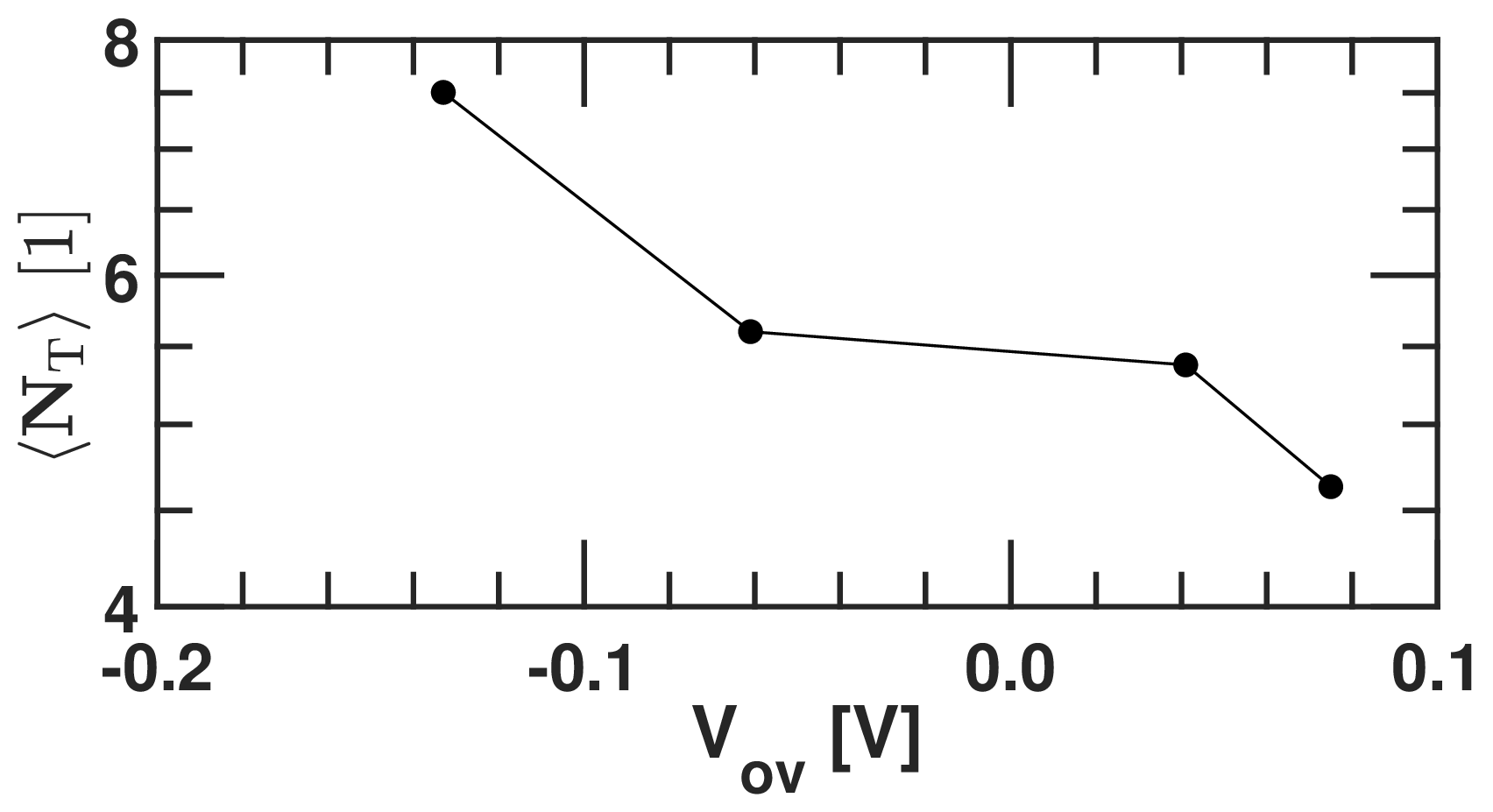}\label{subfig:NtVov}}
\caption{Average number of noise-active traps per nanosheet device ($N_{\mathrm{T}}$) versus (a) $I_{\mathrm{D}}$ and (b) $V_{\mathrm{ov}}$. The $N_{\mathrm{T}}$ is extracted by fitting the distributions shown in Fig.~\ref{fig:probit_Noise} with the defect-centric model shown in Eq.~\ref{eq:DefectCentric} \cite{Kaczer10}.}
\label{fig:Nt_withFit}
\end{figure}

 We measured the drain current noise power spectral density ($S_{\mathrm{id}}$) at different $I_{\mathrm{D}}$ (100 nA, 200 nA, 1 \textmu A, and 2 \textmu A) for each nanosheet. For each device, we computed the root mean square voltage noise at the gate ($V_{\textrm{rms,noise}}$) by integrating the input-referred drain current noise power spectral density ($S_{\mathrm{vg}}=S_{\mathrm{id}}/g_{\mathrm{m}}^2$) as
\begin{equation}
V_{\mathrm{rms,noise}}=\sqrt{\int_{10}^{1000}S_{\mathrm{vg}}\ df}\ \ \ .
\label{eq:Vrms}
\end{equation}
The CDF of \( V_{\mathrm{rms,noise}} \) is shown on a probit scale in Fig.~\ref{fig:probit_Noise} and fitted using the defect-centric model proposed in~\cite{Kaczer10}. This model describes gate-voltage fluctuations due to charge trapping and detrapping in terms of the number of active charged defects \( N_{\mathrm{T}} \), assumed to be Poisson distributed, and the contribution of each individual defect to \( V_{\mathrm{rms,noise}} \), assumed to follow an exponential distribution with mean value \( \eta \). The resulting CDF is given by
$$
H_{\eta,\langle N_{\mathrm{T}} \rangle}(V_{\mathrm{rms,noise}})=
$$
\begin{equation}
\sum_{n=0}^{\infty}
\frac{\mathrm{e}^{-\langle N_{\mathrm{T}} \rangle}\langle N_{\mathrm{T}} \rangle^{n}}{n!}\left[
1 - \frac{n}{\eta}\,
\Gamma\!\left(n, \frac{V_{\mathrm{rms,noise}}}{\eta}\right)
\right].
\label{eq:DefectCentric}
\end{equation}
From the fit, we extract the average number of defects contributing to noise per nanosheet \( \langle N_{\mathrm{T}} \rangle \) as a function of \( I_D \) and gate voltage overdrive ($V_{\mathrm{ov}}=V_{\mathrm{GS}}-V_{\mathrm{T}}$), as shown in Fig.~\ref{fig:Nt_withFit}. The overdrive values in Fig.~\ref{subfig:NtVov} correspond to the average $V_{\mathrm{ov}}$ at each measured $I_{\mathrm{D}}$.
\par
\begin{figure}[!tb]
\centering
\includegraphics[width=0.45\textwidth]{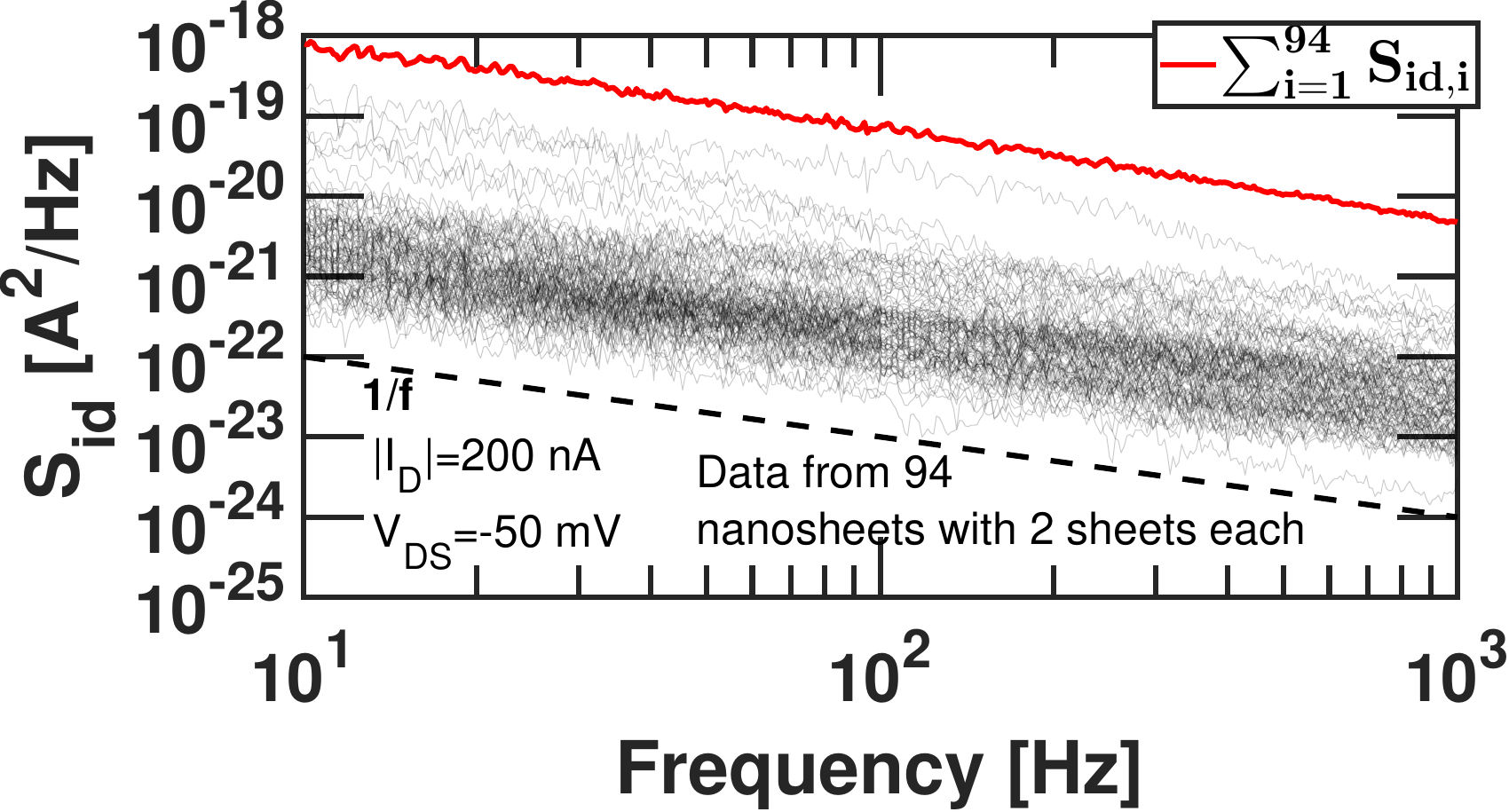}
\caption{$S_{\mathrm{id}}$ versus frequency for all 94 nanosheets pMOSFETs at $|I_{\mathrm{D}}|$=200 nA. The sum of all the spectra exhibits a clear 1/f trend and is highlighted in red.}
\label{fig:SidSum_100nA}
\end{figure}

\begin{figure}[!tb]
\centering
\includegraphics[width=0.45\textwidth]{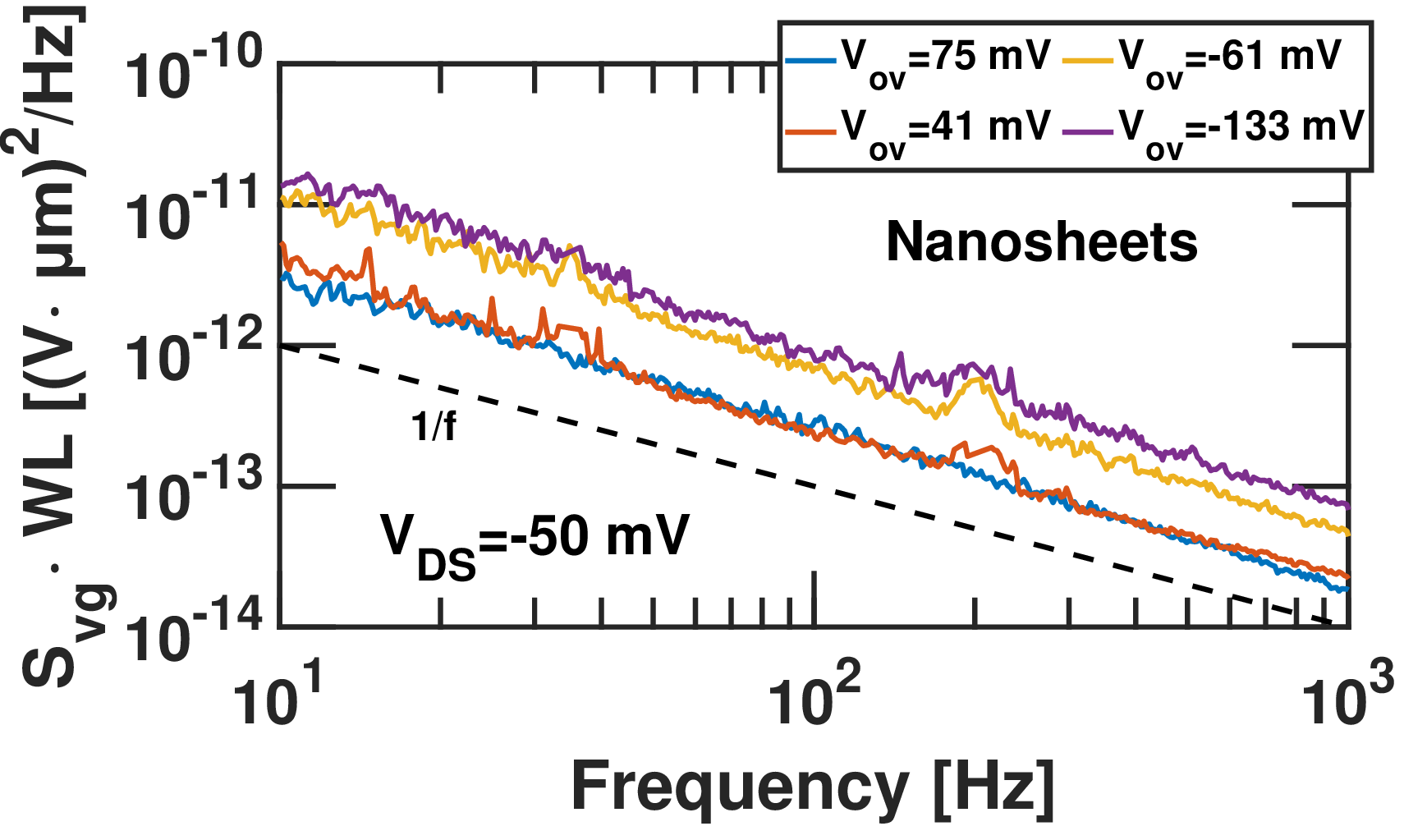}
\caption{Area-normalized $S_{\mathrm{vg}}$ versus frequency at different $V_{\mathrm{ov}}$, obtained by summing spectra from 94 nanosheet pMOSFETs.}
\label{fig:Svg_ALL_Nanosheets}
\end{figure}

\begin{figure}[!tb]
\centering
\includegraphics[width=0.45\textwidth]{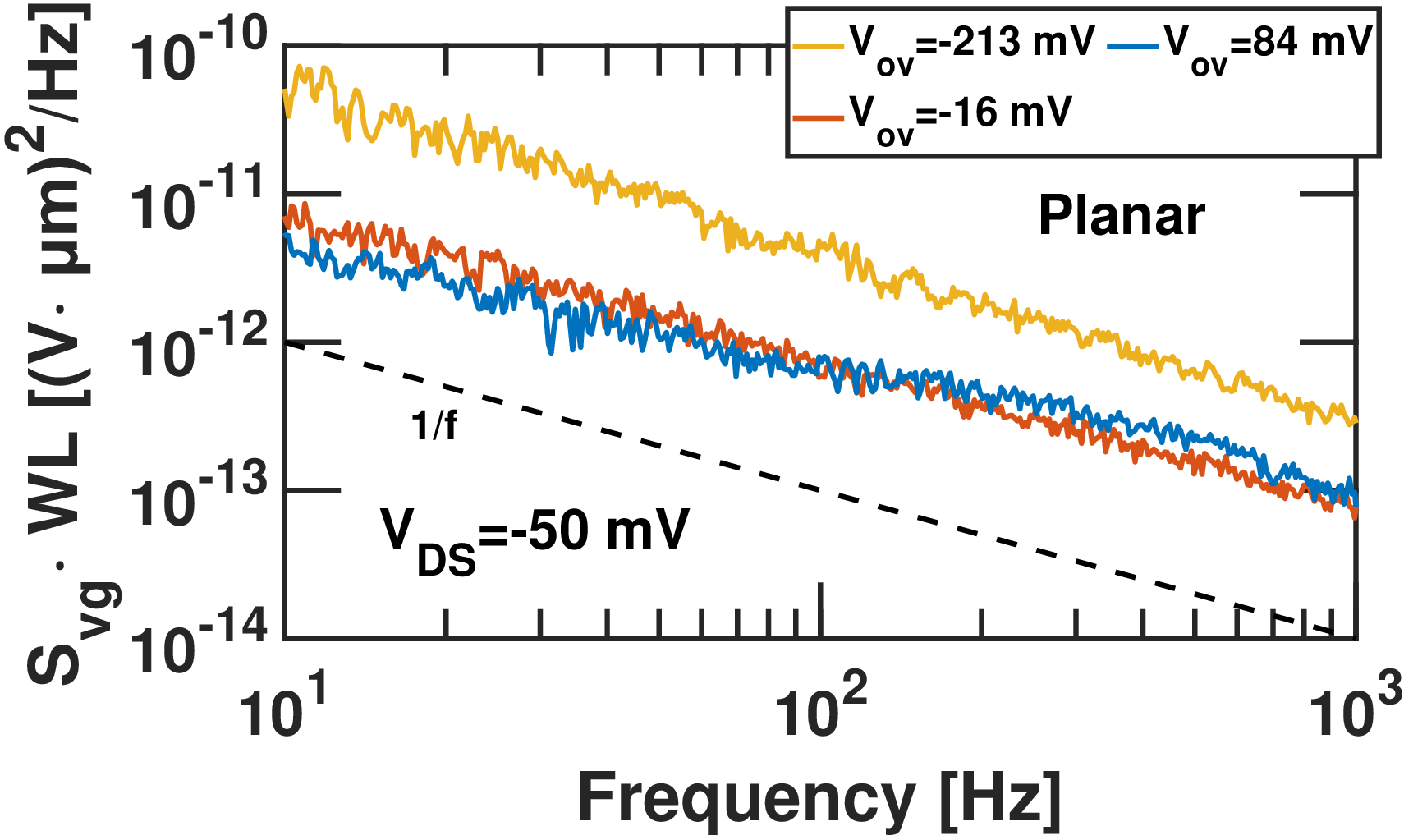}
\caption{Area-normalized $S_{\mathrm{vg}}$ versus frequency at different $V_{\mathrm{ov}}$ for a large-area planar pMOSFET.}
\label{fig:Svg_ALL_Planar}
\end{figure}

 1/f noise represents the average response of many defects fluctuating during the measurement. To obtain a clean 1/f noise spectrum representative of the average nanosheet gate stack quality, we sum the noise spectra from individual devices as if the devices were connected in parallel (Fig.~\ref{fig:SidSum_100nA}). This yields the 1/f noise spectrum of an equivalent large-area device, enabling a representative characterization of the average gate stack quality. The overall $S_{\mathrm{vg}}$ from the summed nanosheet spectra is obtained as
\begin{equation}
S_{\mathrm{vg}}=\frac{\sum_{i=1}^{94}S_{\mathrm{id,i}}}{\sum_{i=1}^{94} g_{\mathrm{m,i}}^2}\ \ \ .
\label{eq:SvgSum}
\end{equation}
We plot the area-normalized $S_{\mathrm{vg}}$ (notice that $S_{\mathrm{vg}}$ scales as 1/$WL$ \cite{Kirton89,Ghibaudo91}) versus gate voltage overdrive $V_{\mathrm{ov}}$ obtained from the sum of the nanosheet noise spectra (Eq.~\ref{eq:SvgSum}) in Fig.~\ref{fig:Svg_ALL_Nanosheets}. The overdrive values in Fig.~\ref{fig:Svg_ALL_Nanosheets} correspond to the average $V_{\mathrm{ov}}$ at each measured $I_{\mathrm{D}}$. The $S_{\mathrm{vg}}$’s in Fig.~\ref{fig:Svg_ALL_Nanosheets} follow a perfect 1/f dependence, as expected when summing up the noise-response of a large ensemble of traps and the noise increase for higher $V_{\mathrm{ov}}$’s. For comparison, Fig.~\ref{fig:Svg_ALL_Planar} plots the $S_{\mathrm{vg}}$ normalized by area of a large area planar pMOSFET ($W{=}L{=}1\ \textrm{\textmu m}$) with the same RMG stack and thermal budget, and no specific anneal to cure dielectric defects. After proper area normalization, nanosheet noise amplitudes are comparable to those of planar devices.
\par
\begin{figure}[!tb]
\centering
\includegraphics[width=0.45\textwidth]{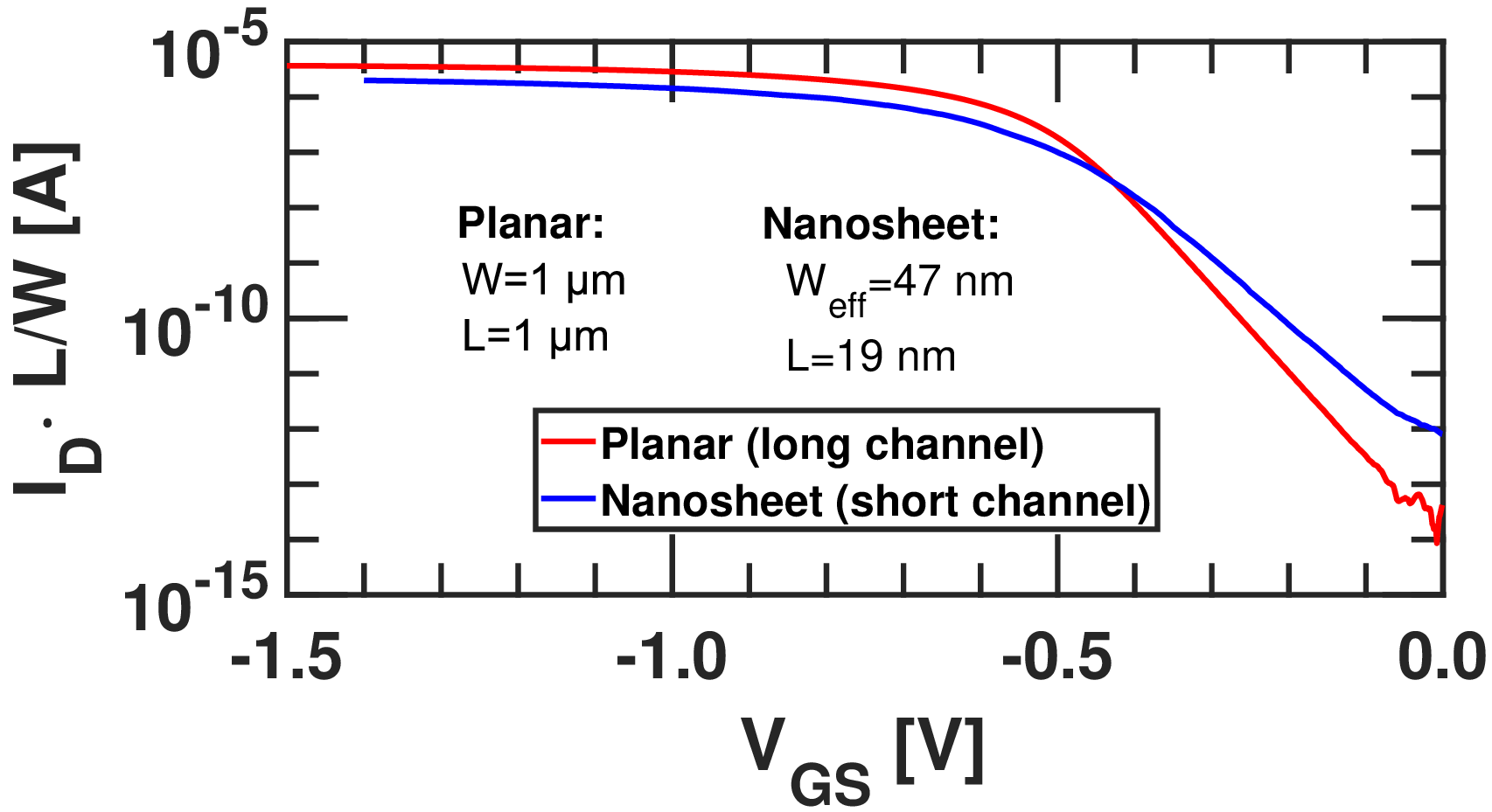}
\caption{Comparison of median $I_{\mathrm{D}}-V_{\mathrm{GS}}$'s (normalized by the aspect ratio $W$/$L$) for planar and nanosheet pMOSFETs. Notice that the SS is higher for the nanosheet due to short-channel effects.}
\label{fig:IdVg_PlanarVsNanosheet}
\end{figure}

\begin{figure}[!tb]
\centering
\includegraphics[width=0.45\textwidth]{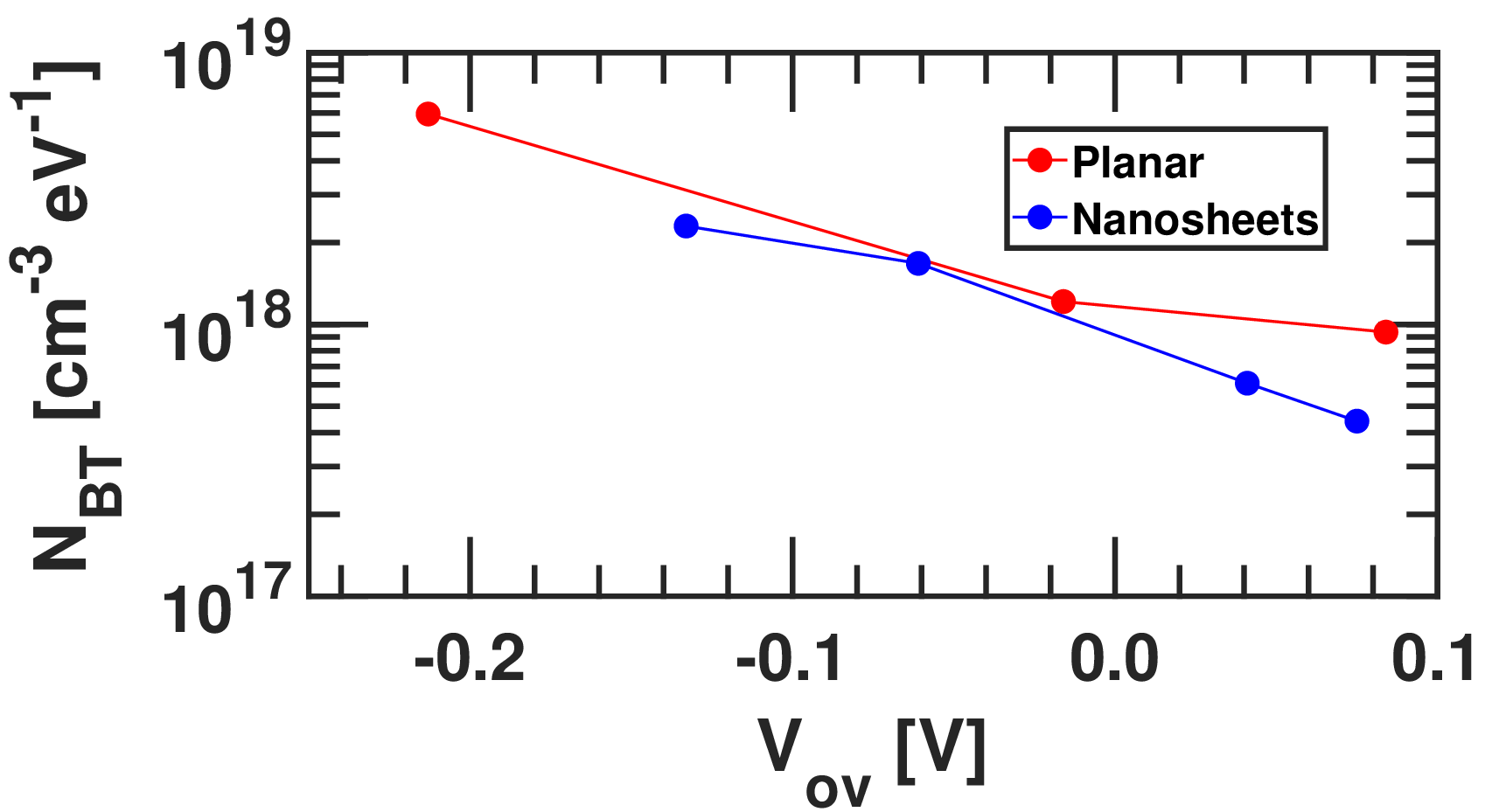}
\caption{Comparison between effective trap density ($N_{\mathrm{BT}}$) extracted from 1/f noise in planar and nanosheet pMOSFETs with the same gate stack and no dedicated dielectric reliability anneals. Notice that the values are consistent between the two architectures. }
\label{fig:NBT_Planar_Vs_Nanosheet}
\end{figure}

\begin{figure}[!tb]
\centering
\includegraphics[width=0.45\textwidth]{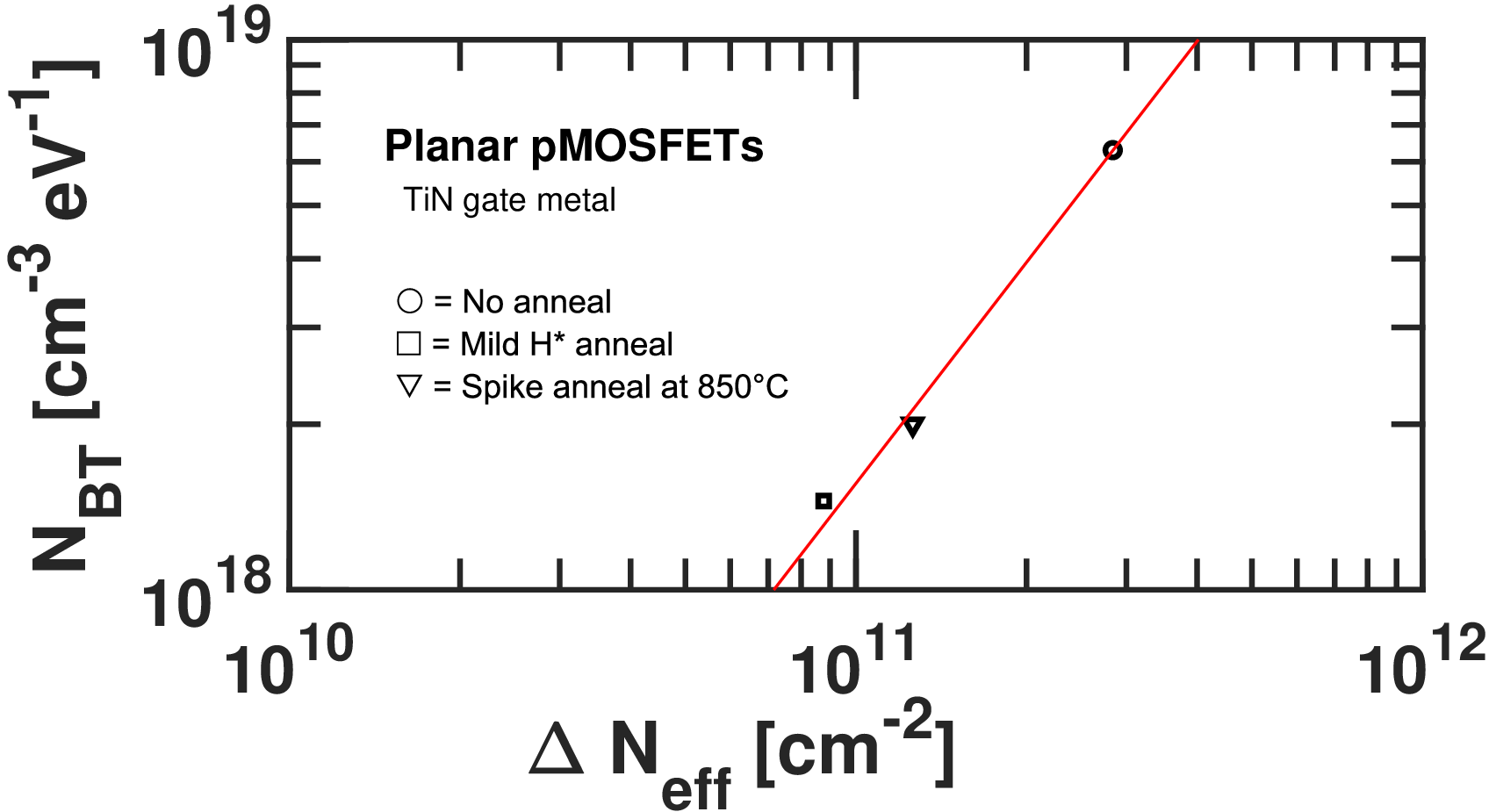}
\caption{Comparison between $\Delta N_{\mathrm{eff}}$ (extracted from BTI) vs $N_{\mathrm{BT}}$ (extracted from 1/f noise) for planar HKMG pMOSFETs \cite{Asanovski2024} highlighting the correlation between the two metrics and the impact of gate stack processing on transistor reliability.}
\label{fig:CorrelationBTINoise_v3}
\end{figure}

 We next compare in detail the characterization of the large planar area pMOSFET with the nanosheet one. First, we compare the median $I_{\mathrm{D}}-V_{\mathrm{GS}}$  in the linear region (Fig.~\ref{fig:IdVg_PlanarVsNanosheet}) and see that the devices exhibit similar $V_\mathrm{T}$, but different SS due to short channel effects. Under the assumption that carrier number fluctuations dominate \cite{Fleetwood15,Vandamme08}, we extract the effective trap density using the relation in \cite{Jayaraman89,Ghibaudo91}
\begin{equation}
N_{\mathrm{BT}}=S_{\mathrm{vg}}\cdot \frac{WLC_{\mathrm{ox}}^2\alpha f}{qkT},
\label{eq:NBT}
\end{equation}
where $C_{\mathrm{ox}}$ is the gate dielectric capacitance per unit area, $q$ is the elementary charge, $\alpha$ is a tunneling coefficient estimated through the Wentzel–Kramer–Brillouin (WKB) approximation, $k$ is the Boltzmann constant, $T$ is the temperature, and $f$ is the frequency.\\
Eq.~\ref{eq:NBT} allows us to extract $N_{\mathrm{BT}}$ and compare directly the quality of the gate stack of the planar and nanosheet architectures. As shown in Fig.~\ref{fig:NBT_Planar_Vs_Nanosheet}, the trap density between the two architectures is comparable, showing that there is no penalty in moving to a scaled architecture. Although the electrostatics and carrier confinement differ between planar and nanosheet devices, the observed agreement in noise behavior indicates that gate stack quality, rather than channel geometry, dominates the 1/f noise response in the studied bias regime. As a result, improvements in gate stack processing are expected to translate directly into enhanced reliability and noise performance in nanosheet devices. This interpretation is further supported by the qualitative agreement between the increase of \(N_{\mathrm{T}}\) with \(V_{\mathrm{ov}}\), as extracted using the defect-centric model (Fig.~\ref{subfig:NtVov}), and the corresponding increase of \(N_{\mathrm{BT}}\) derived from \(1/f\) noise measurements (Fig.~\ref{fig:NBT_Planar_Vs_Nanosheet}).\\
\indent In previous studies, we established a correlation between BTI and 1/f noise over a wide range of gate metal work functions and reliability anneals \cite{Asanovski2024,Asanovski2025}. Fig.~\ref{fig:CorrelationBTINoise_v3} illustrates this relationship for planar pMOSFETs with a TiN gate metal, demonstrating that both noise and BTI are primarily governed by the reliability anneal applied to the gate stack, and therefore by the resulting gate‑stack quality.  The correlation between the noise of nanosheet and planar devices demonstrated here suggests that similar gate stack optimization trends would apply for nanosheet architectures, as reliability improvements are directly related to the gate stack quality and processing. 
\section{Conclusions}
We performed a statistically relevant wafer-level 1/f noise characterization of scaled p-type nanosheet transistors and compared them to planar HKMG devices with identical gate stacks and thermal processing. The extracted effective trap densities are similar across both architectures, indicating that the transition to a gate-all-around does not introduce additional noise sources. These findings reinforce that gate stack quality remains the dominant factor in noise performance and reliability of scaled devices.
\section*{Acknowledgment}
Research work performed within imec’s Core Partner Program. This work has been enabled in part by the NanoIC pilot line. The acquisition and operation are jointly funded by the Chips Joint Undertaking, through the European Union’s Digital Europe (101183266) and Horizon Europe programs (101183277), as well as by the participating states Belgium (Flanders), France, Germany, Finland, Ireland and Romania.
\ifCLASSOPTIONcaptionsoff
  \newpage
\fi

\bibliographystyle{IEEEtran}
\bibliography{IRPS}



%

\end{document}